\documentclass[aps,notitlepage,a4paper,superscriptaddress,11pt]{article}

\usepackage[utf8]{inputenc}
\usepackage[dvips]{color}
\usepackage[toc]{appendix}
\usepackage[hang,flushmargin]{footmisc} 
\usepackage{titlesec,titletoc,ntheorem,authblk,abstract,latexsym,microtype,booktabs,amsmath,cleveref,fancyhdr,wrapfig,array,amsfonts, amssymb,geometry,appendix,cite}
\usepackage{dsfont}

\newcommand{\be}{\begin{equation}}
\newcommand{\ee}{\end{equation}}
\newcommand{\bea}{\begin{eqnarray}}
\newcommand{\eea}{\end{eqnarray}}
\newcommand\blfootnote[1]{%
  \begingroup
  \renewcommand\thefootnote{}\footnote{#1}%
  \addtocounter{footnote}{-1}%
  \endgroup
}

\usepackage{xcolor}
\usepackage{graphicx,lipsum}
\usepackage{caption}
\usepackage{subfigure}
\usepackage{amsmath}
\usepackage{amsfonts}
\usepackage{amssymb}
\usepackage{mathtools}
\numberwithin{equation}{section}
\usepackage{titlesec}

\numberwithin{subcase}{case}
\usepackage{framed}
\allowdisplaybreaks
\usepackage[nottoc]{tocbibind}
\usepackage[free-standing-units]{siunitx}
\usepackage{helvet}
\usepackage{url}
\usepackage{titletoc}
\usepackage{blindtext}
\usepackage[gen]{eurosym}

\DeclareMathOperator{\sech}{sech}
\DeclareMathOperator{\cosech}{cosech}

\title{so(2,1) algebra, local Fermi velocity, and position-dependent mass Dirac equation}
\author[1]{Bijan Bagchi} 
\author[1]{Rahul Ghosh}
\author[2]{Christiane Quesne}
\affil[1]{Physics Department, Shiv Nadar University, Gautam Buddha Nagar, Uttar Pradesh 203207, India}
\affil[2]{Physique Nucl\'{e}aire Th\'{e}orique et Physique Math\'{e}matique, Universit\'{e} Libre de Bruxelles,
Campus de la Plaine CP229, Boulevard du Triomphe, B-1050 Brussels, Belgium}

\begin{document}

\maketitle

\begin{abstract}
Abstract: We investigate the (1+1)-dimensional position-dependent mass Dirac equation within the confines of so(2,1) potential algebra by utilizing the character of a spatial varying Fermi velocity. We examine the combined effects of the two when the Dirac equation is equipped with an external pseudoscalar potential. Solutions of the three cases induced by ${\rm so}(2, 1)$ are explored by profitably making use of a point canonical transformation. 
\end{abstract}

\blfootnote{E-mails: bbagchi123@gmail.com, rg928@snu.edu.in,  Christiane.Quesne@ulb.be}
{Keywords: Dirac equation, position-dependent mass, local Fermi velocity, so(2,1) algebra}

\section{Introduction}

Study of Dirac equation has been of perennial interest in problems of relativistic and non-relativistic quantum mechanics \cite{thal}. It has found numerous applications in many areas of physics including the ones that give physical understanding of the properties of charge carriers of graphene's electronic structure (see, for example, \cite{nov, net, dow2016, gall1, gall2, luo}). 

In recent times, the Dirac equation with a varying mass has received particular attention in the light of wave packet dynamics and effective envelopes
    of wave propagation in topological materials \cite{ber, hu, rag, xie}. The need for a consistent treatment of position-dependent mass (PDM) in an effective Hamiltonian was initially taken up by von Roos \cite{vonroos} while examining the dynamics of free carriers in semiconductors of nonuniform chemical composition. In the ensuing decades, the general interest in PDM problems has gradually grown as is evidenced from the huge amount of literature accumulated in relation to compositionally graded crystals \cite{Gel}, quantum dots \cite{Ser}, liquid crystals \cite{Bar} and other theoretically appealing contexts \cite{alha, ioffe}. In a PDM setting, one has to confront an extended form of the Schr\"{o}dinger equation that depends on a wide range of effective potentials containing different choices of ambiguity parameters \cite{bag2, mus1, ikh, mus2}. The presence of such ambiguity parameters has indeed opened up many pathways for exploration (see, for example, \cite{dut, dha, carinena, cruz1, cruz2, cunha}). In particular, Quesne used extensively the point canonical transformation (PCT) to analyze different variants of systems endowed with PDM \cite{que1, que2, que3, que4}.   

Because of a gap formation in graphene \cite{gui} the need to include a spatially varying Fermi velocity was pointed out by Downing and Portnoi in \cite{dow2017}. Tunneling spectroscopy experiments also confirmed this issue \cite{juan, yan, jang}. From a theoretical side, the necessity of a local Fermi velocity (LFV) was subsequently examined in \cite{mus3, oliv1, oliv2, ghosh}, which included a study on the electronic transport in two-dimensional strained Dirac materials \cite{phan}. Of course, results on the constant Fermi velocity case with respect to the scalar shape-invariant Schr\"odinger Hamiltonians relevant to a class of Dirac-like matrix Hamiltonians exist for the stationary 1-dimensional Dirac equation with pseudoscalar potentials \cite{kuru}.

In this paper we propose to inquire into the working of the combination of PDM and LFV in the (1+1)- dimensional Dirac equation following the framework of an so(2,1) potential algebra. The algebraic use of the corresponding Casimir to facilitate generating exactly solvable potentials is known for a long time due to the ground laying works of Alhassid et al \cite{alh1, alh2, alh3, alh4, alh5} and also of some other ventures who have explored
more general possibilities for the group generators \cite{suk, eng, bq1, bq2, lev}. Additionally,  \cite{bq3} discussed
how such generators are modified when the PDM restriction is imposed upon them. 

The paper is organised as follows. The next section summarizes the basic role of the ${\rm so}(2, 1)$ algebra in a PDM background. We follow up in section 3 by writing down the PDM Dirac equation with LFV enforced and give a mathematical formulation of the scheme in terms of a pair of coupled differential equations corresponding to the wavefunctions embodying a two-component spinor. In section 4, we give the complete classification of the associated pseudoscalar potentials for the three cases induced by ${\rm so}(2, 1)$. Finally, in section 5, some concluding remarks are presented. 
 
\section{so(2,1) algebra in a PDM background}
 
In this section we sketch briefly the results of \cite{bq3}. In the PDM framework the kinetic energy operator $\hat{T}$ is given by \cite{vonroos}

\begin{gather}\label{T}
 \hat{T}= \frac{1}{4} (\mu^\eta(x) \hat{p} \mu^\beta(x) \hat{p} \mu^\gamma(x) + \mu^\gamma(x) \hat{p} \mu^\beta(x) \hat{p} \mu^\eta(x)), \quad \hat{p} = -i \hbar \frac{\partial}{\partial x},
\end{gather}
where $\mu(x)$ is the mass function and the ambiguity parameters $\eta, \beta$ and $\gamma$ are constrained by the relation
\begin{equation}
    \eta+\beta+\gamma = -1
\end{equation}
to ensure Hermiticity of $\hat{T}$. The above representation of $\hat{T}$ is of course not unique but implementation of other choices does not lead to much new physics \cite{Geller}. 

Setting $\mu (x) = \mu_0 M(x)$, where $M(x)$ is a positive dimensionless function, and adopting units $\hbar = 2 \mu_0 = 1$,  the time-independent modified Schr\"{o}dinger equation corresponding to (2.1) acquires the form
\begin{gather} \label{pdmse1}
 H\psi(x)=\Big[- \frac{d}{dx}\frac{1}{M(x)}\frac{d}{dx}+V_{{\rm eff}}(x) \Big]\psi(x)= E\psi(x) 
 \end{gather}
with an associated energy $E$. The effective potential $V_{{\rm eff}}(x)$ depends on $M(x)$ and the given potential field $\mathcal{V}(x)$ in the manner
\begin{gather} \label{veffmain}
V_{{\rm eff}}(x)= \mathcal{V}(x) +\frac{1}{2}(\beta+1)\frac{M''}{M^2}-\left(\eta(\eta+\beta+1)+\beta+1\right)\frac{M'^2}{M^3},
\end{gather}
where the primes correspond to spatial derivatives. Equation (2.3) is in quite general form in that it involves the presence of all the parameters $\eta$, $\beta$, $\gamma$ subject to their obeying (2.2).

 Turning to the employment of the ${\rm so}(2, 1)$ algebra, its signature commutation relations defined in terms of its generators $J_+, J_-, J_0$ are

\begin{equation}
    [J_+, J_-] = -2J_0, \quad [J_0, JTurning_{\pm}] = \pm J_{\pm}
\end{equation}
Furthermore,  an irreducible representation of the potential algebra so(2,1) corresponding to the type $D_k^+$ has basis states that point to the eigenfunctions of different Hamiltonians, having the same energy level. The basis kets $|ks \rangle$ are simultaneous eigenstates of the operators $J_0$ and $J^2$

\begin{equation}
J_0 |ks \rangle = s |ks \rangle, \quad J^2 |ks \rangle = k(k-1) |ks \rangle, \quad s = k, k+1, k+2, \ldots.
\end{equation}

Similar in spirit to the representations of $J_0$ and $J_{\pm}$ formulated by Englefield and Quesne \cite{eng}, the following ones were proposed in \cite{bq3} in the context of PDM that satisfy the algebra (2.5) 

\begin{eqnarray}
&& J_0 = -i \frac{\partial}{\partial \phi}, \\
&& J_{\pm} = e^{\pm i\phi}\left [ \pm \frac{1}{\sqrt{M}}\frac{\partial}{\partial x} + F(x) \left ( i \frac{\partial}{\partial \phi} \mp \frac{1}{2} \right ) + G(x) \right ],
\end{eqnarray}
where $\phi$ is an auxiliary parameter and an appropriate change of variable has been made to bring the generators in one to one correspondence with the constant mass case. Note that the basis kets can be expressed in the form $|ks \rangle = \chi_{ks} (x) e^{is\phi}$. The resulting constraints on $F(x), G(x)$ are related by the coupled equations 

\begin{equation}
    F' = \sqrt{M} (1-F^2), \quad G' = - \sqrt{M} FG.
\end{equation}
 The Casimir $J^2$ is defined by
 
 \begin{equation}
     J^2 = J_0^2 \mp J_0 - J_{\pm} J_{\mp},
 \end{equation}
 which in terms of the representations (2.7) and (2.8) cast the extended Schr\"{o}dinger equation (2.3) in the form
 
 \begin{gather} \label{pdmse2}
 H\chi(x) \equiv \Big[-\frac{1}{\sqrt{M}}\frac{d}{dx}\frac{1}{\sqrt{M}}\frac{d}{dx}+ V_s (x) \Big]\chi(x) = \mathcal{E}_k \chi(x) 
 \end{gather}
 where $\chi (x) \equiv \chi_{ks}(x)$ and $\mathcal{E}_k$ is given by
 
 \begin{equation}
     \mathcal{E}_k = - \left(k - \frac{1}{2}\right)^2, \quad k = 0, 1, 2, \ldots.
 \end{equation}
 In (2.11) the one-parameter family of potentials stands for\footnote{There are some serious misprints in \cite{bq3}. For instance, there is the factor $\frac{1}{\sqrt{M}}$ missing in (9) while the fraction in the coefficient of $F'$ should read $\frac{1}{4}.$}
 
 \begin{equation}
     V_s = \frac{1}{\sqrt{M}} \left [ \left (\frac{1}{4} - s^2 \right )F' + 2 s G' \right] + G^2, \quad s = k, k+1, k+2,\ldots.
 \end{equation}
Thus ${\rm so}(2, 1)$ as a potential algebra defines the above class of potentials in a PDM background induced by the mass function $M(x)$ and conforming to the same set of energy eigenvalues $\mathcal{E}_k$. Observe that the ambiguity parameters remain with $V_{{\rm eff}}$.

Finally, noting that $\chi(x) = \chi_{ks}(x)$ are the eigenfunctions of different Hamiltonians but conform to the same energy level \cite{eng, bq3},  we transform $\chi(x) \rightarrow [M(x)]^{-\frac{1}{4}} \psi(x)$, to rewrite (\ref{pdmse2}) in the manner
\begin{equation} \label{soSE}
\Big[- \frac{d}{dx}\frac{1}{M}\frac{d}{dx}+ \frac{M''}{4M^2}-\frac{7 {M'}^2}{16 M^3}+ V_s (x) \Big] \psi(x) = \mathcal{E}_k \psi(x).
\end{equation} 
(2.14) can be looked upon as an alternative but equivalent equation to (2.11). Observe that in the present scenario the ambiguity parameters are arbitrary and remain with $V_{eff}$. In (2.14), $V_s$ plays the role of the effective potential. This point is exploited later. Various other choices have been explored in the literature which are special cases of the von Roos. These include the BenDaniel-Duke \cite{ben1966} $(\eta = \gamma = 0, \beta = -1)$, Zhu–Kroemer \cite{zhu1983} $(\eta = \gamma = -\frac{1}{2}, \beta = 0)$ and Mustafa–Mazharimousavi \cite{mus1} $(\eta = \gamma = -\frac{1}{4}, \beta = - \frac{1}{2})$ orderings. Actually, the last two are only two physically allowed possibilities to choose $\eta$, $\beta$ and $\gamma$ in (2.1) and that these are the only parametrizations that pass the Dutra-Almeida test \cite{dut} as good orderings.

\section{PDM Dirac equation with LFV}

In the standard form of the Dirac Hamiltonian \cite{jun2020, bg2021}

\begin{equation}\label{H_D1}
  H_D = v_f \sigma_x \hat{p}_x  +\sigma_y W(x)  +\sigma_z  m_0 v_f^2 +\mathds{1} V(x), 
\end{equation}
where $\hat{p}_x=-i \frac{\partial}{\partial x}$, $m_0$ corresponds to a constant mass spin-$\frac{1}{2}$-particle and $v_f$ is the constant Fermi velocity. Other quantities appearing in (3.1) are the electrostatic potential  $V(x)$, the pseudoscalar potential $W(x)$ and the block-diagonal unit matrix $\mathds{1}$, while the Pauli matrices are 
\begin{gather}
\sigma_x = \left( \begin{array}{cc} 0 & 1  \\ 1 & 0  \end{array} \right), \quad \sigma_y = \left( \begin{array}{cc} 0 & -i  \\ i & 0  \end{array} \right), \quad \sigma_z = \left( \begin{array}{cc} 1 & 0  \\ 0 & -1  \end{array} \right). 
\end{gather}
In the following study, we will ignore the effects of $V$ as indeed suggested by the analysis of \cite{jun2020, ish} through the use of intertwining operators. 

The introduction of PDM and the Fermi velocity signified by $m(x)$ and $v_f(x)$, respectively, operating as local variables necessitates a modification of $H_D$. In such a situation, the Dirac Hamiltonian transforms to

\begin{gather} \label{H_D}
  H_D = \sqrt{v_f(x)} \sigma_x \hat{p}_x \sqrt{v_f(x)} +\sigma_y W(x)  +\sigma_z  m(x) v_f^2 (x),
\end{gather}
where $m = m(x)$ and $v_f = v_f (x)$. Put in the two-dimensional matrix form, $H_D$ reads

\begin{gather}\label{H_Dmatrix}
     H_D = \left( \begin{array}{cc} m v_f^2  &  -i \sqrt{v_f} \partial \sqrt{v_f} -iW  \\ -i \sqrt{v_f} \partial \sqrt{v_f}  + i W & - m v_f^2   \end{array} \right).
\end{gather}
When applied on a spinor whose components are $ (\psi_+ \quad \psi_-)^T$, this gives

\begin{gather}
     \left( \begin{array}{cc} m v_f^2  &  -i \sqrt{v_f} \partial \sqrt{v_f} -iW  \\ -i \sqrt{v_f} \partial \sqrt{v_f}  + i W & - m v_f^2   \end{array} \right)  \left(
     \begin{array}{cc} \psi_+  \\ \psi_-  \end{array} \right)  = E  \left(
     \begin{array}{cc} \psi_+  \\ \psi_-  \end{array} \right),
\end{gather}
where $E$ is the energy eigenvalue. Explicitly we have the set of coupled equations  

 \begin{gather}
 (-i \sqrt{v_f} \partial \sqrt{v_f} -i W) \psi_- = D_- \psi_+, \end{gather}
 \begin{gather} \label{lowercomponent}
 (-i \sqrt{v_f} \partial \sqrt{v_f} +i W) \psi_+ = D_+ \psi_- ,
\end{gather}
where the quantities $D_{\pm}$ correspond to $D_{\pm} = E \pm m v_f^2$. When disentangled, the equation for the upper component $\psi_+$ becomes
\begin{gather} \label{GSED1}
-\frac{v_f^2}{D_+} \psi_+'' - \Big(\frac{v_f^2}{D_+}\Big)' \psi_+' + \Big[\frac{1}{D_+} \Big(W^2 -\frac{1}{4}{v'_f}^2-\frac{1}{2}v_f v''_f\Big) + v_f \Big(\frac{W}{D_+}\Big)'     \nonumber \\ 
- \frac{1}{2}v_f v'_f \Big(\frac{1}{D_+}\Big)'  \Big] \psi_+ = D_- \psi_+
\end{gather}
and a similar one holds for $\psi_-$ on elimination of $\psi_+$ from (3.6) and (3.7). In (3.8) the primes refer to the derivatives with respect to the variable $x$.
In the following we will focus on (3.8).

Making use of  Mustafa's constancy condition \cite{mus3} 

\begin{equation} \label{mustafaconstant}
    m(x)v_f^2(x)= A, 
\end{equation}
where $A$ is a positive constant, enables us to get rid of the explicit presence of $m(x)$ in (3.8). In other words, we are led to the equation
\begin{equation} \label{UncoupledDirac1}
\Big[- \frac{d}{dx} v_f^2 \frac{d}{dx} + \Big( W^2 -\frac{1}{4}{v'_f}^2-\frac{1}{2}v_f v''_f + v_f W' \Big) \Big] \psi_{+} = \Big(E^2- A^2 \Big) \psi_{+}.         
\end{equation}
Equation (3.10) is in direct correspondence with the PDM-induced Schr\"{o}dinger equation (\ref{soSE}). The connection 
\begin{gather} \label{relvfmass}
 v_f^2(x) = \frac{1}{M(x)}
\end{gather}
 is obvious. Further, comparing (3.9) and (3.11), $A$ may be interpreted as the ratio of the (physical) Dirac mass and the auxiliary mass $M(x)$. 

Apart from (3.11), the following consistency relations are valid too
\begin{gather} \label{VeffVs}
 W^2 -\frac{1}{4}{v'_f}^2-\frac{1}{2}v_f v''_f + v_f W' =  V_s+\frac{M''}{4M^2}-\frac{7 {M'}^2}{16 M^3}, \\
 E^2 = A^2  - \left(k - \frac{1}{2}\right)^2,  \quad k = 0,1,2,\ldots.
\end{gather}
To get real energies, one must have the criterion $A^2 \geq (k - \frac{1}{2})^2$. $A$ is kept arbitrary\footnote{Assuming $A=1$ would be rather restrictive in the sense that  $E^2$ would be constrained to values $\leq 1$.} but subject to satisfying this condition.

Using (\ref{relvfmass}) we also easily verify
\begin{equation}
    \frac{M''}{4M^2}-\frac{7 {M'}^2}{16 M^3}= -\frac{1}{4}{v'_f}^2-\frac{1}{2}v_f v''_f.
\end{equation}
Therefore the remaining part of the equation (\ref{VeffVs}) can be projected as follows
\begin{gather} \label{psedoscalarpotential}
    W^2 + v_f W' =V_s.
\end{gather}
Equation (\ref{psedoscalarpotential}) is in the Riccati form. This result is new and is central to our present work. A point to note is that in \cite{bag3}, where a study was made in connection with deformed shape invariance condition of supersymmetric quantum mechanics, the role of $v_f(x)$ was played by the deforming function $f(x)$ appearing there.

\section{Classification of pseudoscalar potentials}

It was shown in \cite{bq3} that the sign of the quantity $\omega = \frac{F^2 - 1}{G^2}$ dictates the different choices of $F$ and $G$ that satisfy (2.9). That only three choices could be made for $F$ and $G$ was first pointed out in \cite{eng} in realizing the dynamical potential algebras for certain types of potentials that later found relevance in the contexts of supersymmetric quantum mechanics \cite{kha,fer,jun} and parity-time symmetric theories \cite{bq1, bq2}. The results are summarized below

\begin{eqnarray}
&& \omega = -\frac{1}{b^2} < 0: F(x) = \tanh [u(x) -c], \quad G(x) = b \sech [u(x) -c], \\
&& \omega = 0: F(x) = \pm 1, \quad G(x) = b e^{\mp u(x)}, \\
&& \omega = \frac{1}{b^2} > 0: F(x) = \coth [u(x) -c]. \quad G(x) = b \cosech [u(x) -c], 
\end{eqnarray}
where $b$ is a real constant, and the quantity $u(x)$ appears due to the following PCT

\begin{equation}
    u(x) - c = \int^x \sqrt{M(t)} dt,
\end{equation}
$c$ being a real constant. It is introduced to get rid of the factor $\frac{1}{\sqrt{M}}$ in the expressions of the two generators in (2.8). For more elaboration on the PCT concerning its application side we refer to \cite{que1}. The potentials corresponding to the three types of solutions (4.1) - (4.3) have the same eigenvalues and their  common properties extend to the calculation of the wave functions. Here we point out that the respective wave functions can be determined first by solving the operator relation $J_- \chi_0 e^{ik\phi} =0$ for $\chi_0 \equiv \chi_{kk}$ and then recursively $\chi_n = \chi_{k, k+n}, n = 1, 2,...$, by evaluating $J{_+}^n \chi_0 e^{ik\phi}$. The resulting chain of solutions are \cite{bq3}

\begin{eqnarray}
&& \chi_0 \sim G^{k-\frac{1}{2}} e^{\int \sqrt{M} G dx} \\
&& \chi_1 \sim \left [G - (k-1)F \right ] G^{k-\frac{3}{2}} e^{\int \sqrt{M} G dx} 
\end{eqnarray}
and so on for the higher ones. Unfortunately, an algebraic approach like the present one cannot directly provide the normalization of wavefunctions and the issue needs to be tackled differently \cite{eng}.

We now consider two explicit cases corresponding to $M$ and $v_f$ being constants, and both of them being treated as local quantities.

\subsection{\boldmath $M$ and $v_f$ are constants} 

Setting $s = k > 0$ and assuming\footnote{Taking $M = 1$ would give from (3.8) and (3.9) $m(x) = A$. We can fix $A$ to be the constant mass $m_0$ appearing in the Dirac Hamiltonian in (3.1).} without loss of generality $M=1$,  which implies from (3.11) $v_f = 1$ as well, we derive from (4.1), (4.2) and (4.3) the following solutions for the pseudoscalar potential $W(x)$,

\begin{eqnarray}
&& W = b \sech (x-c), \\
&& W = b e^{-(x-c)}, \\
&& W = b \cosech (x -c),
\end{eqnarray}
In each of the above cases $k = \frac{1}{2}$. These are hyperbolic forms different combinations of which have been encountered in the literature before \cite{nog} while constructing reflectionless pseudoscalar potentials for the Dirac equation. 

\subsection{\boldmath $M$ and $v_f$ are local entities}

(a) Let us deal with the case (4.1) first. Assuming a plausible form

\begin{equation}
    u (x) - c = \tanh^{-1} x
\end{equation}
implies straightforwardly from (4.1) the results

\begin{equation}
    F (x) = x, \quad G(x) = b\sqrt{1 - x^2}, \quad \quad |x| < 1.
\end{equation}
Next, matching (4.4) with (4.8), the outcome is the following solution for $M(x)$ 

\begin{equation}
    M(x) = \frac{1}{(1-x^2)^2}, \quad |x| < 1,
\end{equation}
along with 

\begin{equation}
    v_f (x) = 1-x^2
\end{equation}
from (3.11). Hence from (2.13) we get for $V_s$ at $s = k$

\begin{equation}
  V_s = \left(\frac{1}{4} - k^2 + b^2\right)(1-x^2) - 2kbx \sqrt{1-x^2}, \quad |x| < 1.
\end{equation}
When Eq.\ (4.12) is compared with (3.15), it yields the pseudoscalar potential 
\begin{equation}
    W= b \sqrt{1-x^2}, \quad |x| < 1
\end{equation}
subject to the following restriction
    
\begin{equation}    
 k =\frac{1}{2}.
\end{equation}
As already noted, the potential algebra provides a common platform for the determination of the associated eigenfunctions with the same eigenvalue. Corresponding to (4.11) and (4.12), the wave functions can be read off from (4.5) and (4.6) namely,

\begin{align}
\chi_0 \sim e^{b \sin ^{-1}(x)}  \\ 
\chi_1 \sim \left [ b\sqrt{1-x^2} + \frac{1}{2} x \right ] (1-x^2)^{-\frac{1}{2}} \chi_0
\end{align}
where $|x| < 1$ and so on. The wave functions are all well behaved within this finite range.

Figure 1 shows a sample graph of $V_s$ enclosed within the interval $(-1, 1)$ and assuming positive values of $b$. For small values of the parameter $b$, the curve turns around after crossing the $x$-axis showing two distinct portions, namely the upper and lower, on both sides of it. However, for large $b$, the left portion dominates resembling an inverted oscillator which reaches a maximum, then falls off, goes down the $x$-axis and subsequently curls over.  \\

(b) Taking a typical form of $u(x) = x$, we run into a similar set of solutions as obtained in the case discussed in  section $4.1$. More precisely we get a damping exponential form for the pseuodoscalar potential.\\

(c) To address the case (4.3), we take the choice

\begin{equation}
    u (x) - c = \coth^{-1} x.
\end{equation}
This furnishes the following forms

\begin{equation}
    F (x) = x, \quad G(x) = b\sqrt{x^2 - 1}, \quad \quad |x| > 1.
\end{equation}
As a result the mass function turns out to be

\begin{equation}
    M(x) = \frac{1}{(x^2 - 1)^2}, \quad |x| > 1,
\end{equation}
from (4.4) and (4.15). This implies for the LFV

\begin{equation}
    v_f (x) = x^2 - 1
\end{equation}
using (3.11). From (2.13), we thus arrive at the following class of accompanying ${\rm so}(2, 1)$ potentials at $s = k > 0$

\begin{equation}
  V_s = \left(\frac{1}{4} - k^2 + b^2\right)(x^2 -1) + 2kbx \sqrt{x^2 - 1}. \quad |x| > 1.
\end{equation}
A representative graph of $V_s$ is plotted in Figure 2. Unlike the one of Figure 1, the interval $-1<x<1$ on the $x$-axis is excluded because in this region $V_s$ becomes imaginary. 

As an implication, we obtain by solving $(3.15)$ the accompanying pseudoscalar potential 
\begin{equation}
    W= b \sqrt{x^2 - 1}, \quad |x| > 1,
\end{equation}
subject to the constraint
\begin{equation}    
k =\frac{1}{2}.
\end{equation}
For this case, the wave functions can be worked out to be
\begin{align}
\chi_0 \sim e^{b\cosh^{-1} (x)}, \quad b < 0\\ 
\chi_1 \sim \left [ b\sqrt{x^2 -1} + \frac{1}{2} x \right ] (x^2 -1)^{-\frac{1}{2}} \chi_0, \quad b < 0
\end{align}
where $|x| > 1$ and so on, and we have to impose b to be negative to ensure their convergence behaviour. See that $\chi_0$ is given in terms of inverse hyperbolic cosine whose range is the interval $[1,+\infty)$. The asymptotic behaviour of the remaining wave functions is controlled by $\chi_0$ apart from coefficients which tend to a constant value for $|x| > 1$.

A representative graph of $V_s$ is plotted in Figure 2. Unlike the one of Figure 1, the interval $-1<x<1$ on the $x$-axis is excluded because in this region $V_s$ becomes imaginary. For small values of negative $b$, $V_s$ opens out in two distinct branches as will be clear from the figure. For large values of negative $b$, $V_s$ looks like the potential of a harmonic oscillator with a flat horizental bottom.

\begin{figure}[ht]
  \centering
  \begin{minipage}[b]{0.45\textwidth}
    \includegraphics[width=\textwidth]{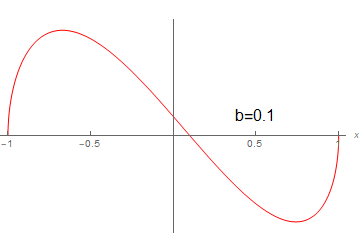}
  \end{minipage}
  \hfill
  \begin{minipage}[b]{0.45\textwidth}
    \includegraphics[width=\textwidth]{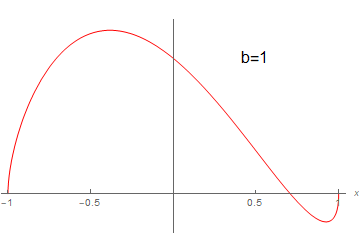}
  \end{minipage}
\caption{Plot of the potential $V_s$ as given by (4.14). 
}
\end{figure} 

\begin{figure}[ht]
  \centering
  \begin{minipage}[b]{0.45\textwidth}
    \includegraphics[width=\textwidth]{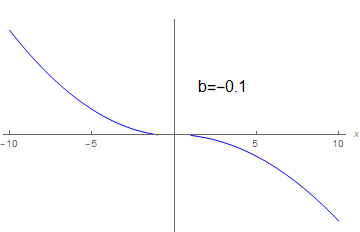}
  \end{minipage}
  \hfill
  \begin{minipage}[b]{0.45\textwidth}
    \includegraphics[width=\textwidth]{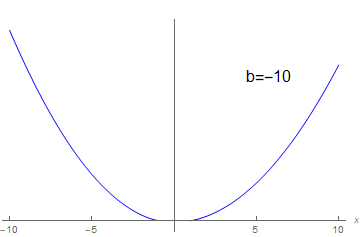}
  \end{minipage}
  \caption{Plot of the potential $V_s$ as given by (4.23). 
  }
\end{figure}

From the foregoing analysis, we see that we encounter three types of analytical solutions for the pseudoscalar potentials, in each case pointing to an inverse square rational functional form of the mass function except when the guiding functions $F(x)$ and $G(x)$ are respectively constant (implying the Fermi velocity to be constant as well) and of exponential types. The mass function shows a singularity at $x = \pm 1$ which is avoided by defining it in the appropriate intervals. Note that $M(x)$ is independent of the ambiguity parameters as defined in $(2.1)$ and symmetrical about $x = \pm 1$ . Evidently, it approaches the unity value asymptotically with respect to $x$. The LFV is accordingly constrained to be a second-degree polynomial. Concerning the solutions of the pseudoscalar potential, we point out that for the two cases (4.1) and (4.3) our results are new and valid in the singularity-free regions $|x|<1$ and $|x|>1$, respectively. In the literature, other forms of the pseudoscalar potential have been studied \cite{ikot} for the solutions of one-dimensional Dirac equation with the pseudoscalar Hartmann potential \cite{hart}. The form is given by $W(x) = -\frac{a}{x} + b \frac{e^{-bx}}{x}$, where $a$ and $b$ stand for the coupling strengths of the one-dimensional Coulomb and Yukawa potentials, but notice that in such a model too the singularity at $x = 0$ is present. Our aforementioned derivation provides a new set of additions in the list.

Finally, we remark on the possibility of deviation from the condition (3.11) when $v_f$ is a constant but $M(x)$ is a varying function of position. This specific scheme was analyzed in detail in an earlier study \cite{bq3} and reflect solutions corresponding to the mass-deformed versions of Scarf II, Morse and generalized P\"{o}schl-Teller potentials.

\section{Concluding remarks}

To conclude, we made a systematic study of (1+1)-dimensional position-dependent mass
Dirac equation in the framework of so(2,1) algebra by additionally taking into account local variation
of the Fermi velocity. We showed that the role of the latter is significant to establish the
consistency of the resulting structure of the eigenvalue problem after disentanglement of
the relevant coupled equations, and the extended Schr\"{o}dinger equation implied by the
Casimir operator of the so(2, 1) algebra. The generated three classes of solutions are shown
to yield the corresponding new forms of pseudoscalar potentials.

\section{Acknowledgment}

We thank the referees for constructive suggestions. One of us (RG) thanks Shiv Nadar University for the grant of senior research fellowship. CQ was supported by the Fonds de la Recherche Scientifique - FNRS under Grant Number 4.45.10.08.

\section{Data availability statement}

All data supporting the findings of this study are included in the article.









\end{document}